\documentclass[]{article}
\usepackage{amsmath}
\usepackage{amsfonts}
\usepackage{amssymb}
\usepackage{graphicx}
\usepackage{amsfonts}
\usepackage[english]{babel}
\usepackage[utf8]{inputenc}
\usepackage{times}
\usepackage[T1]{fontenc}
\usepackage{physics}

\begin{document}

\title{Revisiting Quantum Volume Operator}

\author{Leonid Perlov\\
Department of Physics, University of Massachusetts,  Boston\\
leonid.perlov@umb.edu\\[2ex]
}

\date{ January 10, 2018}

\maketitle

\begin{abstract}
In this paper we introduce the n-dimensional hypersurface quantum volume operator by using the n-dimensional holonomy variation formula. Instead of trying to construct the n-dimensional hypersurface volume operator by using the n-1 dimensional hypersufrace volume operators, as it is usually done in 3d case, we introduce the n-dimensional volume operator directly. We use two facts - first, that the area of the n-dimensional hypersurface of the n+1 dimensional manifold $\cite{Thiemann}$ is the volume of the n dimensional induced metric and secondly that the holonomy variation formula $\cite{Lewandowski}$ is valid for the n-dimensional hypersufrace in the n+1 manifold with connection values in any Lie algebra.

\end{abstract}

\section{Introduction}
The area of the n-dimensional hypersurface of the n+1 dimensional manifold $\cite{Thiemann}$ is the volume of the n dimensional induced metric and can be written as:
\begin{equation}
A(S) = \int_{U} d^{n}u \sqrt{n_a(u) n_b(u)E^a_jE^b_j}
\end{equation}
, where $a, b, j = 1, \mbox{..}, (n+1)$\\
%
%
The main goal of this paper is to introduce the quantum volume (area) operator for the volume of the n-dimensional hypersurface in n+1 dimensional space. Instead of trying to create the n-dimensional volume operator by  using n-1 dimensional area electric flux operators as it is usually done in LQG $\cite{RovelliBook}, \cite{Thiemann}$ we propose to introduce the volume operator directly. We will illustrate first the idea on the 3 dimensional volume of the 4 dimensional space first before moving to the n-dimensional hypersurface. \\[2ex]
The paper is organized as follows. In the next section \ref{sec:3DVolumeOperator} we derive the 3d volume operator by using the new approach. In section $\ref{sec:NDimVolumeOperator}$ we generalize this approach for the volume operator of the n-dimensional hypersurface. The discussion section $\ref{sec:Discussion}$ concludes the paper. 

\section{ 3D Volume Operator }
\label{sec:3DVolumeOperator}

Let us consider the 4-dimensional Lorentzian manifold with the Ashtekar complex  variables $\cite{Ashtekar}$ : self-dual connection $A^{IJ}_{\mu}$ with the values in $sl(2,C) \otimes C = sl(2,C) \oplus sl(2,C)$ and the corresponding conjugate momentum variable - complex valued fluxes ${\Pi}^{\mu}_{IJ}$, where $I, J, \mu = 1,\mbox{...} 4$.  We consider a curve ${\gamma}$ intersecting the 3d hypersurface of the 4 dimensional manifold. The variation of the holonomy along this curve with respect to the connection according to $\cite{Lewandowski}$ will have the following form: \\
\begin{equation}
\label{variation}
 \frac{\delta}{\delta A^{IJ}_{\alpha}(x)} U(A, \gamma) = \int ds \; \dot{\gamma}^a(s) \delta^4 (\gamma(s), x) [U(A, \gamma_1) X_{IJ} U(A, \gamma_2)]
\end{equation}
, where $U(A, \gamma)$ is a holonomy along the path $\gamma(s)$, $X_{IJ}$ are $sl(2,C)$ algebra generators, $I, J = 1, \mbox{..}4$.
The electric field quantum operator inserts $sl(2,C)$ algebra generator $X_{IJ}$ when the curve $\gamma(s)$ intersects the 3d hypersurface. We then introduce the volume grasp operator: 
\begin{equation}
\label{grasp}
\hat{\Pi}_{IJ} U(A, \gamma) = -i\hbar \int_{\Sigma} d \sigma^1 d \sigma^2 d \sigma^3 n_{\alpha}(\vec{\sigma})\frac{\delta}{\delta A^{IJ}_{\alpha}(x(\vec{\sigma}))}
\end{equation}
, where $ n_{\alpha}(\vec{\sigma}) $ is a 3-D hypersurface norm covector
\begin{equation}
n_a =  \epsilon_{\alpha \beta \gamma \delta} \frac{\partial x^{\beta}}{\partial \sigma^1}\frac{\partial x^{\gamma}}{\partial \sigma^2}\frac{\partial x^{\delta}}{\partial \sigma^3}
\end{equation}
By substituting ($\ref{variation}$) into ($\ref{grasp}$) we obtain:
\begin{multline}
\label{grasp1}
\hat{\Pi}_{IJ} U(A, \gamma) = -i\hbar \int_{\gamma} ds \int_{\Sigma} d \sigma^1 d \sigma^2 d \sigma^3 \epsilon_{\alpha \beta \gamma \delta} \frac{\partial x^{\beta}}{\partial \sigma^1}\frac{\partial x^{\gamma}}{\partial \sigma^2}\frac{\partial x^{\delta}}{\partial \sigma^3} \frac{\partial x^{\alpha}}{\partial s} \times
\\ \delta^4 (\gamma(s), x) [U(A, \gamma_1) X_{IJ} U(A, \gamma_2)]
\end{multline}
 If we assume that the 4-dimensional manifold can be spanned by the curves $\gamma$ and 3d hypersurfaces then we obtain:
\begin{equation}
\label{GraspOperator}
\hat{\Pi}_{IJ}U(A, \gamma) = \sum_{p \in ( \Sigma \cap \gamma )} \pm i\hbar \; U(A, \gamma_1^p) \;  X_{IJ} \; U(A, \gamma_2^p) 
\end{equation}
We introduce the operator:
\begin{equation}
\hat{\Pi}_{IJ}^2(V) = \sum_{IJ} \delta^{IJ} \hat{\Pi}_{IJ}(V)\hat{\Pi}_{IJ}(V)
\end{equation}
The operator acts by inserting the algebra generators $X_I$, therefore:
\begin{equation}
\label{E2}
\hat{\Pi}_{IJ}^2(V) \ket{V} = -\hbar^2 \sum_{IJ} \delta^{IJ} (X_{IJ})(X^{IJ}) \ket{V}
\end{equation}
The  $\sum_{IJ} \delta^{IJ} (X_{IJ})(X^{IJ})$ is SL(2,C) Casimir. 
Using the principal series $SL(2,C)$ representation with parameters $(n \in Z, \rho \in C)$ that includes also unitary representations (for $\rho \in R$) and all non-unitary ones (for $\rho \in C$ ) we know that the sum is equal to $\cite{Ruhl}, \cite{Naimark}$, \cite{Knapp}:
\begin{equation}
\label{X2}
\sum_{IJ}  \delta^{IJ}  (X_{IJ})(X^{IJ}) \ket{V} = -\frac{1}{2}(n^2 - \rho^2 - 4)  \ket{V}
\end{equation}
from ($\ref{E2}$) and ($\ref{X2}$) we obtain:
\begin{equation}
\label{E2-2}
\hat{\Pi}_{IJ}^2(V)  \ket{V} = \frac{\hbar^2}{2}(n^2 - \rho^2-4) \ket{V}
\end{equation}
The volume operator is then defined as:
\begin{equation}
\label{Area}
\hat{V} (V) = \lim_{k \rightarrow \infty}  \sum_k \sqrt{\hat{\Pi}_{IJ}^2 (V_k)} =  \lim_{k \rightarrow \infty}  \hbar \sum_k \sqrt{\frac{1}{2}(n^2 - \rho^2-4)}
\end{equation}

\section{ N-Dim Volume Operator }
\label{sec:NDimVolumeOperator}
The same formalism can be generalized to the volume of the n-dimensional hypersurface in the n+1 dimensional space. The difference is n-dimensional connection variables  $A_{\mu}^{IJ}$ with the values in some Lie algebra  $so(n)$ or even in the algebra of non-compact group $sl(n, C)$ with the corresponding n-dimensional conjugate momentum.
We would again consider a curve $\gamma$ intersecting the n-dimensional hypersuface in the n+1-dimensional manifold. 
The the holonomy variation formula is valid for the n+1 dimensional manifold and any connection as mentioned in $\cite{Lewandowski}$: 
\begin{equation}
\label{n-variation}
 \frac{\delta}{\delta A^{IJ}_{\alpha}(x)} U(A, \gamma) = \int ds \; \dot{\gamma}^a(s) \delta^{n+1} (\gamma(s), x) [U(A, \gamma_1) X_{IJ} U(A, \gamma_2)]
\end{equation}
,where $X_{IJ}$ are the generators of the connection $A_{\mu}^{IJ}$ algebra.
By introducing the n-volume grasp operator: 
\begin{equation}
\label{n-grasp}
\hat{\Pi}_{IJ} U(A, \gamma) = -i\hbar \int_{\Sigma} d \sigma^1 d \sigma^2 \mbox{...} d \sigma^n n_{\alpha}(\vec{\sigma})\frac{\delta}{\delta A^{IJ}_{\alpha}(x(\vec{\sigma}))}
\end{equation}
, where $ n_{\alpha}(\vec{\sigma}) $ is the n-dimensional hypersurface norm covector
\begin{equation}
n_a =  \epsilon_{\alpha \beta \gamma \mbox{...}\delta} \frac{\partial x^{\beta}}{\partial \sigma^1}\frac{\partial x^{\gamma}}{\partial \sigma^2}\mbox{...}\frac{\partial x^{\delta}}{\partial \sigma^n}
\end{equation}
By substituting ($\ref{n-variation}$) into ($\ref{n-grasp}$) we obtain:
\begin{multline}
\label{n-grasp1}
\hat{\Pi}_{IJ} U(A, \gamma) = -i\hbar \int_{\gamma} ds \int_{\Sigma} d \sigma^1 d \sigma^2 \mbox{...} d \sigma^n \epsilon_{\alpha \beta  \gamma \mbox{...} \delta} \frac{\partial x^{\beta}}{\partial \sigma^1}\frac{\partial x^{\gamma}}{\partial \sigma^2}\mbox{...}  \frac{\partial x^{\delta}}{\partial \sigma^n} \frac{\partial x^{\alpha}}{\partial s} \times
\\ \delta^{n+1} (\gamma(s), x) [U(A, \gamma_1) X_{IJ} U(A, \gamma_2)]
\end{multline}
by doing integration as in 4-dimensional case we obtain:
\begin{equation}
\label{GraspOperator}
\hat{\Pi}_{IJ}U(A, \gamma) = \sum_{p \in ( \Sigma \cap \gamma )} \pm i\hbar \; U(A, \gamma_1^p) \;  X_{IJ} \; U(A, \gamma_2^p) 
\end{equation}
The volume operator of the n-dimensional hypersurface is then defined similarly to the 3-dimensional case:
\begin{equation}
\label{n-Area}
\hat{V} (V) = \lim_{k \rightarrow \infty}  \sum_k \sqrt{\hat{\Pi}_{IJ}^2 (V_k)} =  \lim_{k \rightarrow \infty}  \hbar \sum_k \sqrt{-C(X_{IJ})}
\end{equation}
, where $C(X_{IJ})$ is the Casimir of the connection algebra spanned by the algebra generators $A_{IJ}$.
Thus for the two dimensional spacelike hypersurface of the 3 dimensional space the volume (area) operator will have the form:
\begin{equation}
\label{2-Areaspacelike}
\hat{V} (V) =   \lim_{k \rightarrow \infty}\hbar \sum_k \sqrt{-C(su(2))} =  \lim_{k \rightarrow \infty} \hbar \sum_k \sqrt{j(j+1)}
\end{equation}
for the two dimensional timelike hypersufrace:
\begin{equation}
\label{2-Areatimelike}
\hat{V} (V) =   \lim_{k \rightarrow \infty} \hbar \sum_k \sqrt{-C(sl(2, R))}
\end{equation}
the three dimensional hypersurface volume operator is:
\begin{equation}
\label{2-Areatimelike}
\hat{V} (V) =   \lim_{k \rightarrow \infty} \hbar \sum_k \sqrt{-C(sl(2, C))} = \lim_{k \rightarrow \infty} \hbar\sum_k \sqrt{\frac{1}{2}(n^2 - \rho^2-4)}
\end{equation}

\section{ Discussion }
\label{sec:Discussion}
In this paper we have suggested a new approach to the n-dimensional hypersurface quantum volume operator. Instead of deriving the 3d volume operator by using the 2-dimensional case grasp operators, we derived it directly by using the holonomy variation formula $\cite{Lewandowski}$. We also generalize this approach for the n-dimensional hypersufrace of the $n+1$ dimensional manifold, with the connection values in any Lie algebra. \\[3ex]
Acknowledgment:
I would like to specially thank Michael Bukatin for multiple fruitful and challenging discussions.

\end{document}